\input harvmac\skip0=\baselineskip
\def\snot{\not\!\!}

% approximately less than

% approximately greater than

%\BergshoeffTU
\lref\BergshoeffTU{ E.~Bergshoeff and P.~K.~Townsend, ``Super
D-branes,'' Nucl.\ Phys.\ B {\bf 490}, 145 (1997)
[arXiv:hep-th/9611173].
%%CITATION = HEP-TH 9611173;%%
}

%\BergshoeffKR
\lref\BergshoeffKR{ E.~Bergshoeff, R.~Kallosh, T.~Ortin and
G.~Papadopoulos, ``Kappa-symmetry, supersymmetry and intersecting
branes,'' Nucl.\ Phys.\ B {\bf 502}, 149 (1997)
[arXiv:hep-th/9705040].
%%CITATION = HEP-TH 9705040;%%
} \lref\dewit{B.~de Wit, P.~G.~Lauwers and A.~Van Proeyen,
``Lagrangians Of N=2 Supergravity - Matter Systems,'' Nucl.\
Phys.\ B {\bf 255}, 569 (1985).}

\lref\cremmer{E.~Cremmer, C.~Kounnas, A.~Van Proeyen, J.~P.~Derendinger,
S.~Ferrara, B.~de Wit and L.~Girardello, ``Vector Multiplets Coupled To
N=2 Supergravity: Superhiggs Effect, Flat Potentials And Geometric
Structure,'' Nucl.\ Phys.\ B {\bf 250}, 385 (1985).}

%\OoguriZV
\lref\OoguriZV{ H.~Ooguri, A.~Strominger and C.~Vafa, ``Black hole
attractors and the topological string,'' arXiv:hep-th/0405146.
%%CITATION = HEP-TH 0405146;%%
}

\lref\strominger{A.~Strominger,
``Special Geometry,''
Commun.\ Math.\ Phys.\  {\bf 133}, 163 (1990).}

\lref\castellani{L.~Castellani, R.~D'Auria and S.~Ferrara,
``Special Geometry Without Special Coordinates,''
Class.\ Quant.\ Grav.\  {\bf 7}, 1767 (1990).}

\lref\ceresole{A.~Ceresole, R.~D'Auria, S.~Ferrara and A.~Van Proeyen,
``Duality transformations in supersymmetric Yang-Mills theories coupled to supergravity,''
Nucl.\ Phys.\ B {\bf 444}, 92 (1995)
[arXiv:hep-th/9502072].}

\lref\hosono{S.~Hosono, A.~Klemm, S.~Theisen and S.~T.~Yau,
``Mirror symmetry, mirror map and applications to complete
intersection Calabi-Yau spaces,''
Nucl.\ Phys.\ B {\bf 433}, 501 (1995)
[arXiv:hep-th/9406055].}

\lref\candelas{P.~Candelas, X.~C.~De La Ossa, P.~S.~Green and
L.~Parkes,
``A Pair Of Calabi-Yau Manifolds As An Exactly Soluble Superconformal
Theory,''
Nucl.\ Phys.\ B {\bf 359}, 21 (1991).}

\lref\ferrara{S.~Ferrara, R.~Kallosh and A.~Strominger,
``N=2 extremal black holes,''
Phys.\ Rev.\ D {\bf 52}, 5412 (1995)
[arXiv:hep-th/9508072].}

\lref\stromicro{A.~Strominger, ``Macroscopic Entropy of $N=2$
Extremal Black Holes,'' Phys.\ Lett.\ B {\bf 383}, 39 (1996)
[arXiv:hep-th/9602111].}

\lref\behrndt{K.~Behrndt, G.~Lopes Cardoso, B.~de Wit, R.~Kallosh,
D.~Lust and
T.~Mohaupt,
``Classical and quantum N = 2 supersymmetric black holes,''
Nucl.\ Phys.\ B {\bf 488}, 236 (1997)
[arXiv:hep-th/9610105].}

\lref\shmakova{M.~Shmakova,
``Calabi-Yau black holes,''
Phys.\ Rev.\ D {\bf 56}, 540 (1997)
[arXiv:hep-th/9612076].}

\lref\alonso{N.~Alonso-Alberca, E.~Lozano-Tellechea and T.~Ortin,
``Geometric construction of Killing spinors and supersymmetry algebras in
%homogeneous spacetimes,''
Class.\ Quant.\ Grav.\  {\bf 19}, 6009 (2002)
[arXiv:hep-th/0208158].} \lref\ssty{A. Simons, A. Strominger, D.
Thompson and X. Yin, to appear.}

\lref\susyvar{ I. Bena and R. Roiban, ``Supersymemtric
pp-wave solutions with 28 and 24 supercharges'', Phys.\ Rev.\
D{\bf 67}, 125014 (2003)
[arXiv:hep-th/0206195]. }

\lref\spinor{T. Mohaupt, ``Black Hole Entropy, Special Geometry
and Strings", hep-th/0007195.}

\lref\beckers{K.~Becker, M.~Becker, and A.~Strominger, ``Fivebranes,
Membranes, and Non-Perturbative String Theory",hep-th/9507158.}

\lref\bpsb{ M. Marino, R. Minasian, G. Moore and A. Strominger,
``Nonlinear Instantons from Supersymmetric $p$-Branes",
hep-th/9911206. }

\lref\msw{ J. Maldacena, A. Strominger and E. Witten, ``Black Hole
Entropy in M-Theory", hep-th/9711053. }

\lref\killing{N. Alonso-Alberca, E. Lozano-Tellechea and T. Ortin,
``Geometric Construction of Killing Spinors and Supersymmetry
Algebra in Homogeneous Spacetimes", hep-th/0208158.}

\lref\poincare{ H. L\"u, C.N. Pope and J. Rahmfeld, ``A Construction
of Killing Spinors on $S^n$'', hep-th/9805151. }

\lref\dzero{ M. Bill\'o, S. Cacciatori, F. Denef, P. Fr\'e,
A. Van Proeyen and D. Zanon, ``The 0-brane action in a general D=4 supergravity
background'', hep-th/9902100. }
%\OoguriZV
\lref\osv{ H.~Ooguri, A.~Strominger and C.~Vafa, ``Black hole
attractors and the topological string,'' arXiv:hep-th/0405146.
%%CITATION = HEP-TH 0405146;%%
}
\lref\gqkahler{ N. Reshetikhin and L. Takhtajan, ``Deformation
Quantization of K\"ahler Manifolds'', math.QA/9907171. }

\lref\kthy{ R. Minasian and G. Moore, ``K-theory
and Ramond-Ramond charge'', hep-th/9710230. }

\lref\kpgus{M.~Cederwall, A.~von Gussich, B.~E.~W.~Nilsson, and
A.~Westerberg,``The Dirichlet super-three-brane in type IIB
supergravity",hep-th/9610148.}

\lref\kpjohn{M.~Aganagic, C.~Popescu, and J.~H.~Schwarz, ``D-brane actions
with local kappa symmetry", hep-th/9610249.}

%\draft 
\Title{\vbox{\baselineskip12pt\hbox{hep-th/0406121}
%\hbox{HUTP-04/XXXX} 
}}{ Supersymmetric Branes in $AdS_2\times
S^2\times CY_3$ }

\centerline{Aaron Simons, Andrew Strominger, David Thompson and Xi
Yin}\smallskip \centerline{ Jefferson Physical
Laboratory}\centerline{ Harvard University}\centerline{ Cambridge,
MA 02138} \vskip .6in \centerline{\bf Abstract} {The problem of
finding supersymmetric brane configurations in the near-horizon
attractor geometry of a Calabi-Yau black hole with
magnetic-electric charges $(p^I,q_I)$ is considered. Half-BPS
configurations, which are static for some choice of global $AdS_2$
coordinate,  are found for wrapped brane configurations with
essentially any four-dimensional charges $(u^I,v_I)$. Half-BPS
multibrane configurations can also be found for any collection of
wrapped branes provided they all have the same sign for the
symplectic inner product $p^Iv_I-u^Iq_I$ of their charges with the
black hole charges. This contrasts with the Minkowski problem for
which a mutually preserved supersymmetry requires alignment of all
the charge vectors. The radial position of the branes in global
$AdS_2$ is determined by the phase of their central charge.  }
\vskip .3in

\smallskip
\Date{June 14, 2004}

\listtoc
\writetoc

%\vfill \eject

\newsec{Introduction}

$AdS_2\times S^2\times CY_3$ flux compactifications of string
theory arise as the near-horizon geometries of type IIA black
holes.  The fluxes are determined from the black hole charges. The
vector moduli of the Calabi-Yau threefold and the radius of the
$AdS_2\times S^2$ are also determined in terms of these charges
via the attractor equations \refs{\ferrara,\stromicro}. These
compactifications are interesting for several reasons. A central
unsolved problem in string theory is to find - assuming it exists
- a holographically dual $CFT_1$ for these
compactifications.\foot{For some cases a dual $CFT_2$ is known
\msw.} Moreover recently a simple and unexpected connection was
found between the partition function of the black hole and the
topological string on the corresponding attractor Calabi-Yau \osv.
In this paper we will further our understanding of these
compactifications by analyzing the problem of supersymmetric brane
configurations.

Following some review in section 2, in section 3 the problem of
supersymmetric branes is analyzed from the viewpoint of the four
dimensional effective ${\cal N}=2$ theory on $AdS_2\times S^2$. This
analysis is facilitated by the recent construction \dzero\ of the
$\kappa$-symmetric superparticle action carrying general electric
and magnetic charges $(u^I,v_I)$ in such theories. It is found
that there is always a supersymmetric trajectory whose position is
determined by the phase of the central charge $Z(u^I,v_I)$. In
global $AdS_2$ coordinates \eqn\glmt{ ds^2 = R^2(-\cosh^2\chi
d\tau^2 + d\chi^2 + d\theta^2+\sin^2\theta d\phi^2) } the
supersymmetric trajectory is at \eqn\ffg{\tanh \chi ={{\rm Re}Z
\over |Z|}.} For the general case $\chi \neq 0$ this
trajectory is accelerated by the electromagnetic forces.  We
further consider $n$-particle configurations with differing charges
and differing central charges $Z_i,~~i=1,..n$, constrained only by
the condition that they all have the same sign for $ {\rm Re}Z_i$.
Surprisingly if the positions of the charges are each determined
by \ffg, a common supersymmetry is preserved for the entire
multiparticle configuration. This is quite different than the case
of fluxless Calabi-Yau-Minkowski compactifications, where there is
a common supersymmetry only if the charges are aligned.
Supersymmetry preservation is possible only because of the
enhanced near-horizon superconformal group.  This phenomena should
have a counterpart in higher AdS spaces and may be of interest for
braneworld scenarios.

In section 4 we consider the problem from the ten-dimensional
perspective. For simplicity we consider only the $AdS_2\times
S^2\times CY_3$ geometries arising from $D0-D4$ Calabi-Yau black
holes. Adapting the analysis of \bpsb\ to this context, we allow
the wrapped branes to induce lower brane charges by turning on
worldvolume field strengths. We will find that there are no
static, supersymmetric D0-branes in global coordinates because
they want to accelerate off to the boundary of $AdS_2$ (there are
static BPS configurations in Poincar\'e coordinates) . For a
D2-brane embedded holomorphically in the Calabi-Yau, we will find
that it is half BPS and sits at $\chi = \tanh^{-1} (\sin
\beta_{CY})$. Here, $\beta_{CY}$ is related to the amount of
magnetic flux on the worldvolume. All D2-brane that are static
with respect to a common global time in $AdS_2$ preserve the same
set of half of the supersymmetries regardless of $\beta_{CY}$.
Similar conclusions hold for D4, D6-branes wrapped on the
Calabi-Yau. We also consider a  D2-brane wrapped on the $S^2$ of
the $AdS_2\times S^2$ product and find that it is once again half
BPS and sits at $\chi = \tanh^{-1} (\sin\beta_{S^2})$.

A related problem is the case of supersymmetric multi-D0-brane configurations
which generate higher brane charges via the Myers effect. This
will be considered in a companion paper \ssty.

\newsec{Preliminaries}
In this section we briefly review some material which will be
needed for our analysis.

 We are interested in type IIA string theory
compactified on a Calabi-Yau 3-fold $M$, with 2-cycles labeled by
$\alpha^A$, where $A=1,2,\cdots,n\equiv h_{11}$. The low energy
effective theory is ${\cal N}=2$ supergravity coupled to $n$
vector multiplets (and also $h_{21}+1$ hypermultiplets which are
not relevant in our discussion).  This theory can be described
using special geometry
\refs{\dewit\cremmer\strominger\castellani{--}\ceresole} and here
we will follow the notation of \refs{\dewit}. The scalar
components of the vector multiplets are described in terms of
projective coordinates $X^I$, $I=0,1,\cdots,n$. The prepotential
${ F}(X^I)$ is holomorphic and homogeneous of degree 2 in
the $X^I$'s. In the large volume limit ${ F}$ is of the form
%\refs{\hosono, \candelas}
\eqn\b{{ F}= D_{ABC}{X^A X^B X^C\over X^0} + \cdots} where
$D_{ABC}=-{ 1\over 6} C_{ABC}$, $C_{ABC}$ being the triple
intersection number of the 4-cycles dual to $\alpha^{A}$, which we
denote by $\Sigma_{A}$.

Extremal black holes of magnetic and electric charge
$(p^0=0,p^A,q_0,q_A)$ are realized as a D4-brane wrapped on 4-cycle
$P=\sum p^A\Sigma_A$ bound with $q_0$ D0-branes, together with $q_A$
gauge field fluxes through the 2-cycles $\alpha^A$. The asymptotic
values of the moduli fields $X^I,F_I\equiv
\partial_I{ F}$ at infinity can be arbitrary. However at the
black hole horizon they approach the fixed point values determined
from the ``attractor equations" \refs{\ferrara,\stromicro}
\eqn\c{p^I = {\rm Re}\,CX^I,~~~ q_I = {\rm Re}\,CF_I.} Using the
tree level prepotential \b, the fixed points of the moduli are
\refs{\behrndt, \shmakova} \eqn\fix{ CX^0=i\sqrt{D\over \hat
q_0},~~~ CX^A = p^A + {i\over 6}\sqrt{D\over \hat q_0}D^{AB}q_B }
where \eqn\defs{ \eqalign{ &D \equiv D_{ABC}p^Ap^Bp^C, \cr &\hat
q_0\equiv q_0+{1\over 12}D^{AB}q_Aq_B, \cr &D_{AB}\equiv
D_{ABC}p^C, \cr & D^{AB}D_{BC}=\delta_C^A. } }

%The horizon area is given by
%\eqn\area{A_H = \pi i|C|^2 (X^I\bar F_I - \bar X^I F_I) = 8\pi
%\sqrt{D \hat q_0}\cdot (l_{4}^{pl})^2 } where $l_4^{pl}$ is the
%4-dimensional Plank length,
%$$l_4^{pl}=\left({\kappa_{10}^2\over 8\pi V_M}\right)^{1/2}=
%{2\sqrt{2} \pi^3 g_s
%\alpha'^2\over {V_M}^{1/2}} $$ where $V_M=(2\pi\sqrt{\alpha'})^6\sqrt{\hat q_0^3/D}$
%is the volume of the CY. The radius of the black hole is
%\eqn\rabh{ R= \sqrt{2}(D q_0)^{1/4} l_4^{pl}
%= {1\over2}g_s\sqrt{\alpha'D\over q_0} }

The near horizon geometry of the 4D extremal black hole is
$AdS_2\times S^2$ with the moduli at their attractor values. We
are interested in string theory on the global $AdS_2\times
S^2\times M$ geometry. The radius $R$ of $AdS_2$ and $S^2$, which
is the same as the radius of the extremal black hole, is
determined in terms of the charges $(p^I,q_I)$ via \eqn\radiusch{
R = \sqrt{2}\left( D \hat q_0\right)^{1\over4}  } where
hereafter we work mainly in four-dimensional Planck units.
% if necessary, in the case $B=0$, the four-dimensional
%Plank length is given by
%$$
%l_4 = \sqrt{8\pi^6g_s^2\alpha'^4\over {\rm vol}(M)}
%$$
%

The metric on the Poincar\'e patch of $AdS_2\times S^2$ is
\eqn\poincpc{ ds^2 = R^2({-dt^2+d\sigma^2\over \sigma^2} +
d\theta^2 + \sin^2\theta d\phi^2) } while the metric is
 \eqn\glmt{ ds^2 =
R^2(-\cosh^2\chi d\tau^2 + d\chi^2 + d\theta^2+\sin^2\theta
d\phi^2) } in global coordinates. In much of the paper we deal
with the case $q_A=0$, and here the RR field strengths are
\eqn\rrsimp{ \eqalign{ F_{(2)} = {1\over R} \omega_{AdS_2},~~~~
F_{(4)} = {1\over R}\omega_{S^2}\wedge J, } } where
$\omega_{AdS_2}=R^2\cosh\chi d\tau\wedge d\chi$ is the volume form
on $AdS_2$, $\omega_{S^2}=R^2\sin\theta d\theta\wedge d\phi$ is
the volume form on the $S^2$, and $J$ is the K\"ahler form on the
Calabi-Yau. In particular, the K\"ahler volume of the 2-cycles $\alpha^A$
are determined by the charges as
\eqn\volumch{ {1\over 2\pi\alpha'}\int_{\alpha^A} J
= 2\pi p^A \sqrt{q_0\over D} }

%It will sometimes be convenient to
%write the 10D spinors corresponding to the $(1,1)$ supersymmetries in type IIA
%supergravity as a ``doublet'' $(\varepsilon_1, \varepsilon_2)$, where
%$\varepsilon_1$ and $\varepsilon_2$ are chiral and anti-chiral 10D Majorana-Weyl
%spinors respectively. In this notation all gamma matrices $\Gamma_M$
%should be replaced by $\Gamma_M\sigma^1$. When we decompose a 10D
%spinor into $SO(3,1)\times SO(6)$ spinors, the gamma matrices decompose
%according to
%\eqn\gammdec{ \eqalign{ & \Gamma_\mu = \gamma_\mu\otimes 1,~~~\mu=0,\cdots,3, \cr
%& \Gamma_m = \gamma_5\otimes \gamma_m,~~~m=4,\cdots,9. } }

\newsec{Four-dimensional analysis}

Flux compactifications on a Calabi-Yau threefold are described by
an effective  $d=4$, ${\cal N}=2$ supergravity  with an
$AdS_2\times S^2$  vacuum solution whose moduli are at the attractor
point with charges $(p^I, q_I)$. This theory contains
zerobranes\foot{We use the term zerobrane in a general sense and
do not specifically refer here to a ten-dimensional D0-brane.}
with essentially arbitrary charges $(u^I,v_I)$ arising from
various wrapped brane configurations. The $\kappa$-symmetric
worldline action of these zerobranes was determined in \dzero. In
this section we use the results of \dzero\ to determine the
possible supersymmetric worldline trajectories.

 The Killing spinor
equation is \eqn\klfd{ \nabla_\mu \epsilon_A - {i\over 2}
\epsilon_{AB} T^-_{\mu\nu} \gamma^\nu \epsilon^B = 0, } where
$\epsilon^A$, $\epsilon_A=(\epsilon^A)^*$ ($A=1,2$) are chiral and
anti-chiral R-symmetry doublets of spinors.
$T^-$ is the anti-self-dual part of the graviphoton
field strength, satisfying \eqn\zgph{ Z_{BH} = {1\over 4\pi }
\int_{S^2} T^- = e^{-{\cal K}/2}\left(F_I p^I - X^I q_I\right) .}
where ${\cal K}=-\ln i(\overline X^I F_I - X^I
\overline  F_I)$ is the K\"ahler potential.
Define the phase of the central charge
$e^{i\alpha}=Z_{BH}/|Z_{BH}|$. Then we can write $T^-=-{i}
e^{i\alpha}(1+i*)F$, where $F={1\over R}\omega_{AdS}$. In terms of
the the doublet of spinors $(\epsilon_1,\epsilon^2)$ and
$(\epsilon^1, \epsilon_2)$, the Killing spinor equation can be
written as \eqn\klspdblt{ \nabla_\mu \epsilon + {i\over
2}e^{-i\alpha\gamma_5}{\not\!\!F}\gamma_\mu \sigma^2\epsilon = 0 .}
Note that there is an ambiguity in choosing the overall phase of
the moduli fields and the central charge,
\eqn\reph{ X^I \to e^{i\theta}X^I, ~~F_I\to e^{i\theta}F_I, ~~
\epsilon\to e^{{i\over 2}\theta\gamma_5}\epsilon ,} so we are free
to set $\alpha=0$.

The solutions to the Killing spinor equation in global
$AdS_2\times S^2$ coordinates \glmt\ are \refs{\alonso}
\eqn\kspsofd{ \eqalign{ \epsilon = &\exp\left({-{i\over 2} \chi
\gamma^0 \sigma^2}\right) \exp\left({{i\over 2} \tau \gamma^1
\sigma^2}\right)R(\theta,\phi) \epsilon_0 \cr
&R(\theta,\phi)\equiv \exp\left({-{i\over 2} (\theta-{\pi/
2})\gamma^{012} \sigma^2}\right) \exp\left({-{i\over 2} \phi
\gamma^{013} \sigma^2}\right)} } where $\epsilon_0$ is a doublet
of arbitrary constant spinors.  Alternatively, in the Poincare
metric \poincpc, the Killing spinors are \poincare\ \eqn\kspptch{
\epsilon = \sigma^{-1/2}R(\theta,\phi)\epsilon_0^+~~~{\rm and}~~~
\epsilon = (\sigma^{1/2} +i\sigma^{-1/2}t\gamma^1 \sigma^2)
R(\theta,\phi)\epsilon_0^-, } where $\epsilon_0^\pm$ are constant
spinors satisfying $-i\gamma^0 \sigma^2 \epsilon_0^\pm =\pm
\epsilon_0^\pm$, and $R(\theta,\phi)$ denotes the rotation on the
$S^2$ as in \kspsofd. Note that $\gamma^\mu$ are the {\sl normalized}
gamma matrices in the corresponding frame.

 The zerobrane action constructed in \dzero\ has a local
$\kappa$-symmetry parameterized by a four-dimensional spinor
doublet $\kappa_A$ on the worldline. In addition the spacetime
supersymmetries $\epsilon_A$ act non-linearly in Goldstone mode on
the worldline fermions.  In general \beckers, a brane
configuration trajectory will preserve a spacetime supersymmetry
generated by $\epsilon$ if  the action on the worldvolume fermions
can be compensated for by a $\kappa$ transformation. This
condition can typically be written \eqn\fdt{(1-\Gamma)\epsilon=0}
where $\Gamma$ is a matrix appearing in the
$\kappa$-transformations. For the case at hand it follows from the
results of \dzero\ that the condition is \eqn\kappro{ \eqalign{
\epsilon_A + e^{i\varphi}\Gamma_{(0)}\epsilon_{AB}\epsilon^B=0 \cr
\epsilon^A + e^{-i\varphi} \Gamma_{(0)}\epsilon^{AB}\epsilon_B =0
} } where $\Gamma_{(0)}$ is the gamma matrix projected to  the
zerobrane worldline, and $e^{i\varphi}$ is the phase of the
central charge $Z$ of the zerobrane, \eqn\ctrz{Z = e^{-{\cal
K}/2}\left(u^I
F_I-v_IX^I\right)=e^{i\varphi}|Z|,} where $(u^I,v_I)$ are its magnetic
and electric charges. In terms of the spinor doublet, one can
write \kappro\ as \eqn\kapdblt{ -i e^{-i\varphi\gamma_5}
\Gamma_{(0)}\sigma^2\epsilon = \epsilon .} Let us solve the
condition for \kapdblt\ to hold along the world line of a
zerobrane sitting at constant $(\chi,\theta,\phi)$. Writing the
Killing spinor as \eqn\epep{ \epsilon = \exp\left({-{i\over 2} \chi
\gamma^0 \sigma^2}\right) \exp\left({{i\over 2} \tau \gamma^1
\sigma^2}\right) \epsilon_0' } where
$\epsilon_0'=R(\theta,\phi)\epsilon_0$, it suffices to solve
\eqn\teqep{ \eqalign{ & -i e^{-i\varphi\gamma_5}
\gamma^0\sigma^2\exp\left({-{i\over 2} \chi \gamma^0
\sigma^2}\right) \epsilon_0' = \exp\left({-{i\over 2} \chi \gamma^0
\sigma^2}\right) \epsilon_0' \cr & -i e^{-i\varphi\gamma_5}
\gamma^0\sigma^2\exp\left({-{i\over 2} \chi \gamma^0
\sigma^2}\right) { \gamma^1 \sigma^2}\epsilon_0' =
\exp\left({-{i\over 2} \chi \gamma^0 \sigma^2}\right) { \gamma^1
\sigma^2}\epsilon_0' } .} Some straightforward algebra simplifies
the above equations to \eqn\epsimp{ \eqalign{ & -i\gamma^0\sigma^2
\left( \cos\varphi + i\cosh\chi \sin\varphi \gamma_5 + \sinh\chi
\sin\varphi \gamma_5 \gamma^0\sigma^2
\right)\epsilon_0'=\epsilon_0' \cr & i \gamma^0\sigma^2 \left(
\cos\varphi - i\cosh\chi \sin\varphi \gamma_5 + \sinh\chi
\sin\varphi \gamma_5 \gamma^0\sigma^2
\right)\epsilon_0'=\epsilon_0' } .} A solution exists only when
\eqn\crit{ \tanh\chi = \cos\varphi ,}
and therefore $\cosh\chi \sin\varphi = \pm1$. Correspondingly
the constraints on $\epsilon_0'$ become
\eqn\epcons{ \eqalign{ \gamma_5 \gamma^0\sigma^2
\epsilon_0'=\mp\epsilon_0' } ,} where the sign on the RHS depends
on the sign of $\sin \varphi$.
This may be written as a condition on $\epsilon_0$,
\eqn\sdep{ \left( 1 \pm e^{{i \over 2} \phi \gamma^{013} \sigma^2} e^{i
(\theta - \pi/2) \gamma^{012} \sigma^2} e^{{i \over 2} \phi \gamma^{013}
\sigma^2} \gamma_5 \gamma^0 \sigma^2 \right) \epsilon_0 = 0\, , }
which makes it clear that zerobranes sitting at antipodal points on the
$S^2$ will preserve opposite halves of the spacetime supersymmetries.

We conclude that a zerobrane
following its charged geodesic in $AdS_2\times S^2$ is half BPS.
 The ``extremal'' case
$\varphi=0$ and $\pi$ corresponds to the probe zerobrane with its
charge aligned or anti-aligned with the charge of the original
black hole. They cannot be stationary with respect to global time
in the $AdS_2$. Using the Killing spinors on the Poincar\'e patch
\kspptch, it is clear that the ``extremal'' zerobranes following
their charged geodesics (static on the Poincar\'e patch) are also
half BPS. In the special case $\varphi={\pi/ 2}$ in \crit\ the
zerobrane moves along an uncharged geodesic and experiences no
electromagnetic forces . This corresponds to the case when the
zerobrane charge is orthogonal to all the black hole charges.

A somewhat surprising feature is that there are supersymmetric
$multiparticle$ configurations of zerobranes with $unaligned$
charges. All ``positively-charged'' zerobranes with $0<\varphi<\pi$
preserve the same set of half of the supersymmetries, and all
``negatively-charged'' zerobranes with $-\pi<\varphi<0$ preserve the
other set. Using the attractor equations the positive charge
condition can be written in terms of the symplectic product of the
black hole and zerobrane charges as  \eqn\tryi{u^Iq_I-p^Iv_I
>0.} Given an arbitrary collection of zerobranes obeying \tryi\
there is a half BPS configuration with the position of each
trajectory determined in terms of the charges of the zerobrane by
\crit. Of course, such a supersymmetric configuration of particles
with unaligned charges is not possible in the full black hole
geometry prior to taking the near horizon limit. The preserved
supersymmetry is part of the enhanced near-horizon supergroup.

This result is consistent with the expectation from the BPS bound.
The energy of a charged zerobrane sitting at position $\chi$ the
$AdS_2$ is given by \eqn\endz{ H=|Z|\cosh\chi -{{\rm Re}(Z\bar
Z_{BH}) \over |Z_{BH}|} \sinh\chi = |Z| \left( \cosh\chi -
\cos\varphi \sinh\chi \right) .}
where the first term comes from the gravitational warping,
and the second term comes from the coupling to the gauge field
potential.
At the stationary point
$\tanh\chi = \cos\varphi$, the energy of the zerobrane is
\eqn\enzb{ |Z\sin\varphi| = {|{\rm Im}Z\bar Z_{BH}|\over |Z_{BH}|}
.} Therefore, as long as ${\rm Im}(Z\bar Z_{BH})$ is always
positive (or negative), the BPS energy for multiple zerobranes is
additive, in agreement with the supersymmetry analysis above.

%In general the zerobrane sitting at different positions on the $S^2$ will
%preserve different halves of the spacetime supersymmetries.  It is
%straightforward to check that zerobrane sitting at antipodal
%points on the $S^2$ will preserve opposite halves of
%supersymmetries.

\newsec{Ten-dimensional analysis}

In this section we analyze supersymmetric brane configurations
from the point of view of the ten-dimensional IIA theory on
$AdS_2\times S^2\times CY_3 $. For simplicity we  will focus on
specific examples rather than the most general solution.

The extremal black hole in type IIA string theory compactified on
a Calabi-Yau manifold $M$ preserves four supersymmetries. After we
take the near horizon limit, the number of preserved
supersymmetries doubles to eight. We consider a background with
only D0 and D4-brane charges, i.e. $q_A=p^0=0$, so that
according to the attractor equations there is no $B$-field.
The RR field strengths in the resulting $AdS_2\times S^2\times
M_6$ are given as in \rrsimp. As shown in Appendix A, the
ten-dimensional Killing spinor doublet is of the form \eqn\tenks{
\eqalign{ &\varepsilon_1 = \epsilon_1\otimes \eta_+ +
\epsilon^1\otimes \eta_-, \cr &\varepsilon_2 = \epsilon^2\otimes
\eta_+ + \epsilon_2\otimes \eta_-, } } where
$\eta_+,\eta_-=\eta_+^*$ are the chiral and anti-chiral
covariantly constant spinors on $M$; $\epsilon_A=(\epsilon^A)^*$,
$\epsilon^{1,2}$ are four-dimensional chiral spinors satisfying
the four-dimensional Killing spinor equation \eqn\ksfd{ \nabla_\mu
\epsilon_A + {i\over 2}{\not\!\!F}^{(2)}\gamma_\mu
(\sigma^2)_{AB}\epsilon^B = 0. } This is the same equation as
\klspdblt\ with $\alpha=0$, and the solutions are given by
\kspsofd, \kspptch.

We want to find all the BPS configurations of D-branes that are
wrapped on compact portions of our background, and are pointlike
in the $AdS_2$. In order for the D-brane to be supersymmetric, we
only need to check that the $\kappa$-symmetry constraint
\eqn\kappas{ \Gamma\varepsilon=\varepsilon } is satisfied, where
$\varepsilon$ is the Killing spinor corresponding to the unbroken
supersymmetry (more precisely, the pullback onto the brane world
volume). The $\kappa$ projection matrix is given by
\refs{\kpgus,\kpjohn,\BergshoeffTU,\BergshoeffKR} \eqn\kapg{
\eqalign{ &\Gamma = {\sqrt{\det G}\over \sqrt{\det(G+{\cal
F})}}\sum_n {1\over 2^n n!}
\Gamma^{\hat\mu_1\hat\nu_1\cdots\hat\mu_n\hat\nu_n} {\cal
F}_{\hat\mu_1\hat\nu_1}\cdots {\cal
F}_{\hat\mu_n\hat\nu_n}\Gamma_{(10)}^{n+{p-2\over
2}}\Gamma_{(0)}\sigma^1,\cr &\Gamma_{(0)} = {1\over (p+1)!\sqrt{\det
G}}\epsilon^{\hat\mu_0\cdots\hat\mu_{p}}
\Gamma_{\hat\mu_0\cdots\hat\mu_p}. } ,} where the hatted indices
label coordinates on the brane world-volume, $G$ is the pullback
of the spacetime metric, and ${\cal F}=F+f^*(B)$ (the $B$-field is
zero in our discussion). See Appendix A for conventions on 10D
gamma matrices.

Unless otherwise noted we will work in global coordinates \glmt.

\subsec{D0-brane}

For a static D0-brane in global coordinates, we have
$\Gamma_{(0)}=\gamma^0$. The $\kappa$-symmetry matrix is \eqn\aap{
\Gamma = \Gamma_{(10)}\gamma^0 \sigma^1 } Writing the doublet
$\varepsilon$ in terms of the 4-dimensional spinor doublet
$\epsilon$ \eqn\aaq{ \varepsilon = \epsilon \otimes \eta_+ +
\epsilon^* \otimes \eta_-\, , } The matrix $\Gamma$ acts on
$\varepsilon$ as $\gamma^0\sigma^1\sigma^3 =-i\gamma^0\sigma^2$.
The $\kappa$-symmetry constraint \kappas\ becomes \eqn\aar{
(1+i\gamma^0 \sigma^2) \epsilon = 0\, . } Using the explicit
solutions of the Killing spinors in global AdS \kspsofd, we see
that \aar\ cannot be satisfied at all $\tau$, so a D0-brane static
in global AdS can never be BPS. This is of course expected since
the charged geodesic cannot be static in global coordinates. On
the other hand, using \kspptch\ we see that a D0-brane static with
respect to the Poincar\'e time is always half BPS, as expected.

\subsec{D2 wrapped on Calabi-Yau, $F=0$}

Now let us consider a D2-brane wrapped on $M$ and static in global
$AdS_2\times S^2$, without any world-volume gauge fields turned
on. The $\kappa$-symmetry matrix is \eqn\aas{ \Gamma = {1\over
2\sqrt{{\det}' G}}\gamma^0 \epsilon^{\hat a\hat b} \Gamma_{\hat a\hat
b} \sigma^1 } where ${\det}'$ takes the determinant of the
spatial components of the world volume metric.
Acting on $\varepsilon$, we have \eqn\aaw{ \eqalign{
\Gamma_{\hat a\hat b} \varepsilon &=
\partial_{\hat a} X^I \partial_{\hat b} X^J \gamma_{IJ} \varepsilon \cr
&= 2 \partial_{\hat a} X^i \partial_{\hat b} X^{\bar{j}}
\gamma_{i\bar{j}} \varepsilon + \partial_{\hat a} X^i
\partial_{\hat b} X^j \gamma_{ij} \varepsilon + \partial_{\hat a}
X^{\bar{i}} \partial_{\hat b} X^{\bar{j}} \gamma_{\bar{i}\bar{j}}
\varepsilon \cr &= 2\partial_{\hat a} X^i \partial_{\hat b}
X^{\bar{j}} \left(-g_{i\bar{j}} \gamma_{(6)}\right) \varepsilon +
\left({1\over 2}\partial_{\hat a} X^i \partial_{\hat b} X^j
\Omega_{ijk} \epsilon\otimes \gamma^k \eta_- + c.c.\right).} }
Firstly, the $\kappa$-symmetry constraint $\Gamma\varepsilon =
\varepsilon$ implies $\epsilon^{\hat a\hat b} \partial_{\hat a}X^i
\partial_{\hat b}X^j \Omega_{ijk}=0$, which means that the
D2-brane must wrap a holomorphic 2-cycle. It then follows that
$\Gamma$ acts on $\varepsilon$ as $\Gamma\varepsilon =
i\gamma^0\gamma_{(6)}\sigma^1\varepsilon = \gamma_{(4)}
\gamma^0\sigma^2\varepsilon$. Therefore \kappas\ becomes \eqn\aba{
(1 -\gamma_{(4)} \gamma^0 \sigma^2)\epsilon = 0\, . } It is clear
that the wrapped D2-brane sitting at $\chi=0$ in $AdS_2$ is half
BPS. Note that the D2-brane without gauge field flux doesn't feel
any force due to the RR fluxes ($q_A=0$), so its stationary
position is at the center of $AdS_2$.

\subsec{D2 wrapped on Calabi-Yau, $F \neq 0$}

With general worldvolume gauge field strength $F$ turned on, the
matrix $\Gamma$ is \eqn\abb{\eqalign{ \Gamma &= {1\over \sqrt{{\det}'
 (G+F)}} \left(1 +{1\over 2}\Gamma^{\hat a\hat
b}F_{\hat a\hat b}\Gamma_{(10)} \right) \gamma^0 \left({1\over
2}\epsilon^{\hat c\hat d}\Gamma_{\hat c\hat d}\right) \sigma^1 \cr
} } An argument nearly identical to the one given in \bpsb\ shows
that the supersymmetric D2-brane must wrap a holomorphic 2-cycle,
and the gauge flux $F$ satisfies \eqn\mmmscd{ {\sqrt{\det G}\over
\sqrt{\det(G+F)}} (f^*J+iF) = e^{i\beta}{\rm vol}_2
 }
where ${\rm vol}_2$ is the volume form on the D2-brane (which is
just $f^*J$ for a holomorphically wrapped brane), and $\beta$ is a
constant phase determined in terms of the D0-brane charge $2\pi n
= {1\over 2\pi\alpha'}\int F$ via \eqn\dzch{ {\tan\beta\over
2\pi\alpha'}\int J = 2\pi n } If the probe D2-brane is
wrapped on the 2-cycle $[\Sigma_2] = n_A\alpha^A$, then using
\volumch\ we have
\eqn\chrel{ \tan\beta = {n\over n_Ap^A}\sqrt{D\over q_0} } Note
that from \mmmscd\ we have $\cos\beta>0$, since $J$ is positive
when restricted to holomorphic cycles. The $\kappa$-symmetry
condition then becomes \eqn\abd{
(1-e^{-i\beta\gamma_{(4)}}\gamma_{(4)}\gamma^0\sigma^2)\epsilon=0 }
These is identical to \kapdblt\ if we set $\varphi = \beta -
\pi/2$. We can immediately read off the conditions for the static
D2-brane to preserve supersymmetry when it sits at $\theta =
\pi/2$, $\phi = 0$ in the $S^2$: \eqn\abe{ \sin \beta = \tanh
\chi\, , \quad ~\cos\beta =  {\rm sech}\chi\, , ~ \quad (1 -
\gamma_{(4)}\gamma^0 \sigma^2 ) \epsilon_0 = 0\, . } We see that
for general $-\pi/2<\beta<\pi/2$, the D2-brane sits at $\chi =
\tanh^{-1}(\sin\beta)$ and is half BPS. In fact they all preserve
the same half supersymmetries, as discuss in section 3.
Anti-D2-branes with gauge field fluxes wrapped on holomorphic
2-cycles will preserve the other half supersymmetries.

\subsec{Higher dimensional D-branes wrapped on the Calabi-Yau}

Let us consider D4, D6-branes that are wrapped on the Calabi-Yau
and sit at constant position in global $AdS_2\times S^2$. We shall
use a trick \BergshoeffKR\ to write the matrix $\Gamma$ as
\eqn\gmtr{ \Gamma = e^{-A/2} \Gamma_{(10)}^{p-2\over
2}\Gamma_{(0)}e^{A/2}\sigma^1 } where \eqn\adefs{ A = -{1\over
2}Y_{\hat a\hat b}\Gamma^{\hat a\hat b}\Gamma_{(10)} } and
$Y_{\hat a\hat b}$ is an anti-symmetric matrix (analogous to the
phase $\beta$ in the previous subsection), related to the gauge
field strength matrix $F_{\hat a\hat b}$ by \eqn\manf{ F = \tanh Y
} By the same arguments as before, one can show that the BPS
D-branes must wrap holomorphic cycles. Note that $A$ acts on the
Killing spinor $\varepsilon$ as $A\varepsilon = -iY_{\hat a\hat
b}(f^*J)^{\hat a\hat b}\gamma_{(4)}\varepsilon$, and
$\Gamma_{(0)}$ acts as $\gamma^0(i\gamma_{(6)})^{p/2}$ (see
Appendix). Let us define $\beta = -Y_{\hat a\hat b}(f^*J)^{\hat
a\hat b}$. The $\kappa$-symmetry constraint can be written as
\eqn\newfkap{ \Gamma\varepsilon = e^{-i\beta\gamma_{(4)}/2}
\Gamma_{(10)}^{p-2\over 2} \gamma^0 (i\gamma_{(6)})^{p/2}
e^{i\beta\gamma_{(4)}/2}\sigma^1\varepsilon =\varepsilon\,.  } We
can simplify this to \eqn\fullsimpk{
-ie^{-i(\beta-p\pi/2)\gamma_{(4)}} \gamma^0 \sigma^2\epsilon =
\epsilon\,. } This equation indeed agrees with \aar, \abd\ in the
cases $p=0,2$. It is also identical to \kapdblt\ provided we set
$\varphi = \beta-p{\pi/2}$. So we conclude that a general
D$p$-brane ($p$ even) wrapped on a holomorphic cycle in the
Calabi-Yau, possibly with world-volume gauge fields turned on,
static in the $S^2$ and following its charged geodesic in the
$AdS_2$ is half BPS. As in \bpsb\ there is a deformation of the
supersymmetry condition on the worldvolume gauge field $F$.

\subsec{D2 wrapped on $S^2$, $F = 0$}

Now let us turn to D2-branes wrapped on the $S^2$ appearing in the
the $AdS_2\times S^2\times M$ product. The $\kappa$-symmetry
matrix is $\Gamma = \Gamma_{(0)}\sigma^1 =\gamma^{023}\sigma^1$.
\kappas\ can be written as \eqn\abj{ (1-\gamma^{023} \sigma^1)
\epsilon = 0\, . } Defining $R(\theta,\phi)$ to be the
$S^2$-dependent factors in \kspsofd, this condition becomes
\eqn\cabk{ \eqalign{ &(1-\gamma^{023}\sigma^1) \exp\left(-{i\over
2}\chi\gamma^0\sigma^2\right) R(\theta,\phi)\epsilon_0=0, \cr &
(1-\gamma^{023}\sigma^1)\exp\left( -{i\over 2}\chi\gamma^0\sigma^2
\right) \gamma^1\sigma^2 R(\theta,\phi)\epsilon_0=0. } } A little
algebra reduces these to \eqn\abk{ \cosh {\chi\over 2}
(1-\gamma^{023} \sigma^1) R(\theta, \phi) \epsilon_0 = \sinh
{\chi\over 2} (1 + \gamma^{023} \sigma^1) R(\theta,\phi)
\epsilon_0 = 0\, . } The only way to satisfy both equations is to
set $\chi=0$. Since $\gamma^{023}\sigma^1$ commutes with
$R(\theta,\phi)$, we end up with the condition \eqn\abm{
(1-\gamma^{023} \sigma^1) \epsilon_0 = 0 } We conclude that the
D2-brane sitting at the center of AdS and wrapped on the $S^2$ is
half BPS.

\subsec{D2 wrapped on $S^2$, $F \neq 0$}

With gauge field strength $F=f\omega_{S^2}$ turned on, the
$\kappa$-symmetry matrix acts on $\varepsilon$ as
\eqn\abn{\eqalign{ \Gamma \varepsilon &= {\sqrt{\det G}\over
\sqrt{\det(G+F)}} \left( 1 + {1\over 2} \Gamma^{\hat{a}\hat{b}}
F_{\hat{a}\hat{b}} \Gamma_{(10)} \right) \Gamma_{(0)} \sigma^1
\varepsilon \cr &= {1\over \sqrt{1+f^2}} \left( 1 + \gamma^{23} f
\Gamma_{(10)} \right) \gamma^{023} \sigma^1 \varepsilon \cr &=
\exp\left({\beta \gamma^{23} \Gamma_{(10)} }\right)
\gamma^{023}\sigma^1\varepsilon = \gamma^{023}\sigma^1
\exp\left(\beta\gamma^{23} \sigma^3 \right)\varepsilon \, , } }
where $f \equiv \tan \beta$ ($\cos\beta>0$). The condition
\kappas\ then becomes \eqn\kabp{ \eqalign{ & (1-\cos\beta
\gamma^{023}\sigma^1-i\sin\beta\gamma^0\sigma^2) \exp\left(
-{i\over 2}\chi\gamma^0\sigma^2\right) R(\theta,\phi)\epsilon_0 =
0,\cr & (1-\cos\beta
\gamma^{023}\sigma^1+i\sin\beta\gamma^0\sigma^2) \exp\left(
{i\over 2}\chi\gamma^0\sigma^2\right) R(\theta,\phi)\epsilon_0 =
0, } } A little algebra yields \eqn\abp{\eqalign{
&\left(1+\sin\beta \coth \chi\right) \epsilon_0 = 0\, , \cr
&\left(1 + \gamma^{023} \sigma^1 \cot \beta \sinh \chi \right)
\epsilon_0 = 0\, . } } This means that $\sin\beta = -\tanh\chi$.
In particular $\beta$, hence $f$, is constant on the world-volume.
The condition on $\epsilon_0$ becomes \eqn\abr{\eqalign{ &(1 -
\gamma^{023} \sigma^1 )\epsilon_0 = 0\, . } } These D-brane
configurations are again half BPS.

\subsec{D-branes wrapped on $S^2$ and the Calabi-Yau}

In general for a D$p$-branes wrapped on $S^2$ times some
$(p-2)$-cycle in the Calabi-Yau, and static in global $AdS_2$, the
matrix $\Gamma$ is essentially the product of the piece on $S^2$
and the piece on Calabi-Yau, \eqn\fstcy{ \Gamma\varepsilon =
\exp\left(-\beta_{S^2}\gamma^{23}\sigma^3 \right) \exp\left(
-i\beta_{CY}\gamma_{(4)} \right) (i\gamma_{(4)})^{p-2\over 2}
\gamma^{023}\sigma^1\varepsilon } where $\beta_{CY}$ and
$\beta_{S^2}$ are the phases related to the world-volume gauge flux
along the Calabi-Yau and $S^2$ directions as before. Define
$\varphi_{CY}=\beta_{CY}-(p-2)\pi/2$,
$\varphi_{S^2}=\beta_{S^2}+\pi/2$. The $\kappa$-symmetry
constraint can be written as \eqn\kfst{
-i\exp\left(-\varphi_{S^2}\gamma^{23}\sigma^3
-i\varphi_{CY}\gamma_{(4)} \right) \gamma^0\sigma^2\epsilon =
\epsilon } This is equivalent to \eqn\solvfa{ \eqalign{ &  \left[
1+i\exp\left(-\varphi_{S^2}\gamma^{23}\sigma^3
-i\varphi_{CY}\gamma_{(4)} \right) \gamma^0\sigma^2\right]
\exp\left( -{i\over 2}\chi\gamma^0\sigma^2 \right)
R(\theta,\phi)\epsilon_0 = 0, \cr &  \left[
1-i\exp\left(\varphi_{S^2}\gamma^{23}\sigma^3
+i\varphi_{CY}\gamma_{(4)} \right) \gamma^0\sigma^2\right]
\exp\left( {i\over 2}\chi\gamma^0\sigma^2 \right)
R(\theta,\phi)\epsilon_0 = 0. } } A little algebra yields
\eqn\aci{\eqalign{ &\left[ \sinh\chi - \cosh\chi
\cos(\varphi_{S^2}-i\gamma_{(4)}\gamma^{23}\sigma^3 \varphi_{CY})
\right] R(\theta,\phi) \epsilon_0 = 0\,, \cr & \left[ \cosh\chi -
\sinh\chi \cos(\varphi_{S^2}-i\gamma_{(4)}\gamma^{23}\sigma^3
\varphi_{CY})\right.\cr &~~~~~~~~~~ \left.
-\gamma^{023}\sigma^1\sin
(\varphi_{S^2}-i\gamma_{(4)}\gamma^{23}\sigma^3 \varphi_{CY})
\right] R(\theta,\phi)\epsilon_0 = 0\, , }} If $\varphi_{CY}$ and
$\varphi_{S^2}$ are both nonzero, the first equation can be
satisfied only if \eqn\onif{ i\gamma_{(4)}\gamma^{23}\sigma^3
R(\theta,\phi)\epsilon_0 =mR(\theta,\phi)\epsilon_0,~~~~m=\pm1. }
However, since $\gamma_{(4)}\gamma^{23}\sigma^3$ does not commute with
$R(\theta,\phi)$\ at generic points on the $S^2$, \onif\ can
never be satisfied. Therefore such wrapped D-branes cannot be BPS.

If $\varphi_{S^2}=0$, $\varphi_{CY}\not=0$, we have
\eqn\onezx{ \tanh\chi = \cos\varphi_{CY} }
and
\eqn\aef{ (1-\gamma_{(4)}\gamma^0\sigma^2) R(\theta,\phi)\epsilon_0=0 }
However, in this case again $\gamma_{(4)}\gamma^0\sigma^2$ does not commute
with $R(\theta,\phi)$ for generic $(\theta,\phi)$, and hence \aef\ has no solution.

If $\varphi_{S^2}\not=0$, $\varphi_{CY}=0$, we find \eqn\onezz{
\tanh\chi = \cos\varphi_{S^2} } and the second equation in \aci\
becomes \eqn\aee{ (1-\gamma^{023}\sigma^1)\epsilon_0=0 } We see
that such D-branes are half BPS.

So far we have neglected an important subtlety. For D4 or
D6-branes wrapped on $S^2$ times some cycle in the Calabi-Yau, the
RR flux $F_{(4)}$ induces couplings of gauge fields on the brane
world-volume \eqn\fscou{ \eqalign{ & \int_{D4} A\wedge F_{(4)}\, ,
\cr & \int_{D6} A\wedge F\wedge F_{(4)}\, , } } Since
$F_{(4)}={1\over R}\omega_{S^2}\wedge J$, we see that for the
D4-brane wrapped on $S^2\times \Sigma_2$ ($[\Sigma_2] =
n_A\alpha^A$), the RR flux induces an electric charge density on
the brane world-volume, of total charge \eqn\ttch{ Q = {1\over
2\pi g_s}\int_{S^2\times \Sigma_2}F_{(4)} = \sum n_Ap^A } Since
the world-volume is compact, the Gauss law constraint requires the
total charge to vanish. So we cannot wrap only a single D4-brane
on $S^2\times \Sigma$. One must introduce fundamental strings
ending on the brane to cancel the electric charges. We then have
$\sum n_Ap^A$ fundamental strings ending on the D4-brane, and
runoff to the boundary of $AdS$. This is interpreted as a
classical ``baryon'' in the dual CFT.

Similarly for the D6-brane wrapped on $S^2\times \Sigma_4$, one would
have nonzero total electric charge on the world-volume if
$\int_{\Sigma_4} F\wedge J\not=0$. This again corresponds to certain
``baryons'' in the dual CFT. In this case, $\varphi_{CY}\not=0$,
and we saw earlier that such branes are not BPS anyway.

Finally, a D6-brane wrapped on $S^2\times \Sigma_4$ with general
gauge field flux in the $S^2$ is half BPS, as shown in \onezz,
\aee.

\bigskip

\centerline{\bf Acknowledgements}
We would like to thank S. Gukov, H. Ooguri and C. Vafa for valuable discussions.
This work was supported in part by DOE grant DE-FG02-91ER40654.

\appendix{A}{The 10-dimensional Killing spinors}

In order to write a ten-dimensional spinor as the tensor product of
four-dimensional and internal (Calabi-Yau) spinors, it is necessary
to work with a tensor product of Clifford algebras. Let $\Gamma^M$
denote the ten-dimensional Clifford algebra matrices, with $M = 0,
\ldots, 10$, $\mu = 0, \ldots, 3$, and $m = 4, \ldots, 9$.  We can
decompose the $\Gamma^M$ into a tensor product of four and
six-dimensional Clifford matrices, denoted by $\gamma^\mu$ and
$\gamma^m$, as \eqn\aaa{\eqalign{ &\Gamma^\mu = \gamma^\mu \otimes
1, \cr &\Gamma^m = \gamma_{(4)} \otimes \gamma^m\, . } } Using a
mostly-positive metric signature, the following matrices have the
desired properties that they anticommute with the appropriate gamma
matrices and square to one: \eqn\aab{\eqalign{ &\Gamma_{(10)} =
-\Gamma^{0123456789},\cr &\gamma_{(4)} = i \gamma^{0123},\cr
&\gamma_{(6)} = i \gamma^{456789}\, . } } With these sign
conventions, $\Gamma_{(10)}$ decomposes in the desired way as
$\Gamma_{(10)} = \gamma_{(4)} \otimes \gamma_{(6)}$.

As an ansatz for the Killing spinors, we assume they take the form
\eqn\aac{ \varepsilon_1 = \epsilon_1 \otimes \eta_+ + \epsilon^1
\otimes \eta_-\, , \qquad \varepsilon_2 = \epsilon^2 \otimes \eta_+
+ \epsilon_2 \otimes \eta_-\, , } where the $\varepsilon$'s are 10D
Majorana-Weyl spinors, the $\eta$'s are 6D covariantly-constant Weyl
spinors on the Calabi-Yau, and the $\epsilon$'s are 4D Majorana
spinors.  We use chiral notation in which the chirality of the
spinor is denoted by the position of the R-symmetry index.  In
particular, $\epsilon(A) = \epsilon^A + \epsilon_A$ where
$\gamma_{(4)} \epsilon^A = \epsilon^A$ and $\gamma_{(4)} \epsilon_A
= - \epsilon_A$.  Of course, there are no Majorana-Weyl spinors in
$3+1$ dimensions; the four-dimensional chiral projections are
related by $\epsilon_A = {\epsilon^A}^*$.  For
the six-dimensional Weyl spinors, we use the standard notation where
$\gamma_{(6)} \eta_\pm = \pm \eta_\pm$.  Since we will work with
type IIA, the tensor products have been chosen such that the
ten-dimensional spinors are of opposite chirality.  In doublet
notation, \eqn\aad{ \varepsilon = \left(\eqalign{ \varepsilon_1 \cr
\varepsilon_2 }\right) } $\Gamma_{(10)} \varepsilon$ can be written
as $-\sigma^3 \varepsilon$. In addition, the following identities
for the spinors $\eta_\pm$ will be useful: \eqn\aae{\eqalign{
\gamma_{\bar{i}} \eta_+ = 0\, , \quad \gamma_{ijk} \eta_+ =
\Omega_{ijk} \eta_-\, , \quad \gamma_{ij} \eta_+ = {1\over 2}\Omega_{ijk}
\gamma^k \eta_-\, , \quad \gamma_{\bar{i}\bar{j}kl} \eta_+ = \left(
g_{k\bar{j}} g_{l\bar{i}} - g_{k\bar{i}} g_{l\bar{j}} \right)
\eta_+\, , \cr \gamma_{i} \eta_- = 0\, , \quad
\gamma_{\bar{i}\bar{j}\bar{k}} \eta_- =
\Omega_{\bar{i}\bar{j}\bar{k}} \eta_+\, , \quad
\gamma_{\bar{i}\bar{j}} \eta_- = {1\over 2}\Omega_{\bar{i}\bar{j}\bar{k}}
\gamma^{\bar{k}} \eta_+ \, , \quad \gamma_{ij\bar{k}\bar{l}} \eta_-
= \left(g_{\bar{k}j} g_{\bar{l}i} - g_{\bar{k}i} g_{\bar{l}j}
\right) \eta_- \, . } }

Given these ans\"atze, we want to check that the supersymmetry
variations of the background vanish modulo conditions on the
four-dimensional Majorana components of the Killing spinors.  Since
we work only with bosonic backgrounds, we need only check the
variations of dilatino and gravitino.

The supersymmetry variation of the dilatino is \susyvar\
\eqn\aaf{
\delta \lambda = {1\over 2} \left( 3 {\not \!\!F}_{(2)} i \sigma^2 + {\not
\!\!F}_{(4)} \sigma^1\right) \varepsilon\, ,
}
where $F_{(2)} = {1\over R} \omega_{AdS_2}$ and $F_{(4)} = {1\over R}
\omega_{S^2} \wedge J$.
Taking note of the fact that $g^{\bar{i}j}\gamma_{\bar{i}j} \eta_\pm = 3
\gamma_{(6)} \eta_\pm$
and
${\snot \omega}_{S^2} = -i {\snot \omega}_{AdS_2} \gamma_{(4)}$,
we find that
\eqn\aag{
{\snot F}_{(4)} \varepsilon = -3 i {\snot \omega}_{AdS_2} \gamma_{(4)}
\gamma_{(6)} \varepsilon = -3 {\snot F}_{(2)} \sigma^3 \varepsilon\, .
}
As a result, the dilatono variation vanishes automatically.

The gravitino variation is
\eqn\aah{
\delta \psi_M = \nabla_M \varepsilon + {1\over 8} \left( {\snot F}_{(2)}
\Gamma_M i \sigma^2 + {\snot F}_{(4)} \Gamma_M \sigma^1 \right) \varepsilon
= 0\, .
}
When the free index is holomorphic in the
Calabi-Yau, this reduces to the following condition:
\eqn\aai{
\left({\snot F}_{(2)} \gamma_m i\sigma^2 + {\snot F}_{(4)} \gamma_m \sigma^1
\right) \varepsilon = 0\, .
}
Using the fact that
$
g^{i\bar{j}} \gamma_{i\bar{j}} \gamma_m \eta_\pm = \gamma_m \gamma_{(6)}
\eta_\pm\, ,
$
we find that
$
{\snot F}_{(4)}\gamma_m \varepsilon = -{\snot F}_{(2)} \gamma_m \sigma^3
\varepsilon\, .
$
This works similarly for an antiholomorphic index, so the gravitino
variation is identically zero when the free index is in the Calabi-Yau.

When the gravitino equation has its free index in the $AdS_2\times S^2$ space,
the variation becomes
\eqn\aaj{
\delta \psi_\mu
= \left[\nabla_\mu \pm {1\over 8} \gamma_\mu \left( {\snot F}_{(2)} i
\sigma^2 - \sigma^1 {\snot F}_{(4)} \right) \right] \varepsilon = 0\, ,
}
where the $\pm$ is $+$ if $\mu$ is in the $S^2$ and $-$ if $\mu$ is in the
$AdS_2$.
Using the same identity used for the dilatino equation, we get
\eqn\aak{
\delta \psi_\mu = \left[ \nabla_\mu \pm {i\over2} \gamma_\mu \snot
F_{(2)} \sigma^2
\right] \varepsilon = \left[\nabla_\mu + {i\over 2} \snot F_{(2)}
\gamma_\mu \sigma^2 \right] \varepsilon\, .
}
Demanding that the terms linear in $\eta_+$ and linear in
$\eta_-$ must vanish separately, we get the 4D equations
\eqn\killeqa{
\left[\nabla_\mu + {i\over 2} {\snot F}_{(2)} \gamma_\mu \sigma^2 \right]
\epsilon = 0\, ,
}
where $\epsilon = \left( \eqalign{\epsilon_1 \cr \epsilon^2 }\right)$.

It is useful to derive the action of $\Gamma_{(0)}={1\over (p+1)!\sqrt{\det G}}\epsilon^{\hat
\mu_0\cdots\hat\mu_p}\Gamma_{\hat \mu_0\cdots\hat \mu_p}$ on the $\eta_\pm$
which live on the world-volume
of holomorphically wrapped D-branes (see \kapg). For
D0-branes we have simply $\Gamma_{(0)}=\gamma^0$. For D2-branes, we have
\eqn\dbact{ \Gamma_{(0)}\eta_\pm
= \gamma^0 \epsilon^{i\bar j}\gamma_{i\bar j} \eta_\pm
= i\gamma^0\gamma_{(6)}\eta_\pm }
For D4-branes, we have
\eqn\dfact{ \Gamma_{(0)}\eta_\pm
= \gamma^0 {1\over 4} \epsilon^{i\bar j k\bar l}
\gamma_{i\bar j k\bar l} \eta_\pm = -\gamma^0\eta_\pm }
where we used the last column of \aae. Finally for D6-branes, we have
$\Gamma_{(0)} = -i\gamma^0\gamma_{(6)}$ using \aab. These formulae
can be summarized as $\Gamma_{(0)}\varepsilon = \gamma^0
(i\gamma_{(6)})^{p/2}\varepsilon$.

\listrefs

\end